\documentclass[journal]{IEEEtran}
\ifCLASSINFOpdf
\else
\fi
\usepackage{textcomp}
\usepackage{enumitem,kantlipsum}
\usepackage{graphicx}
\usepackage{amsmath}
\usepackage{hyperref}
\usepackage{cleveref}
\usepackage{mathtools}
\usepackage{subcaption}
\usepackage{float}
\usepackage{adjustbox}
\usepackage{lipsum}
\usepackage{threeparttable}
\usepackage{flexisym}
\usepackage{multirow}
\usepackage{dblfloatfix}
\usepackage{amssymb,amsmath,amsthm,enumitem}
\usepackage{caption}
\begin{document}
\setlength{\abovedisplayskip}{3pt}
\setlength{\belowdisplayskip}{3pt}
\title{Two-stage building energy consumption clustering based on temporal and peak demand patterns}
\author{Milad Afzalan, Farrokh~Jazizadeh, and~Hoda~Eldardiry
}

\maketitle
\begin{abstract}
Analyzing smart meter data to understand energy consumption patterns helps utilities and energy providers perform customized demand response operations. Existing energy consumption segmentation techniques use assumptions that could result in reduced quality of clusters in representing their members. We address this limitation by introducing a two-stage clustering method that more accurately captures load shape temporal patterns and peak demands. In the first stage, load shapes are clustered by allowing a large number of clusters to accurately capture variations in energy use patterns and cluster centroids are extracted by accounting for shape misalignments. In the second stage, clusters of similar centroid and power magnitude range are merged by using Dynamic Time Warping. We used three datasets consisting of $\sim$250 households ($\sim$15000 profiles) to demonstrate the performance improvement, compared to baseline methods, and discuss the impact on energy management.
\end{abstract}
\IEEEpeerreviewmaketitle

\section{Introduction}
\linespread{0.9} 
\IEEEPARstart{D}{emand} response (DR) mechanisms help electricity providers maintain distribution reliability, reduce generation cost and environmental concerns, and increase utilization of renewables. Residential sector accounts for $\sim$40\% of electricity consumption in the U.S.~\cite{eia1}, making it a proper candidate for DR engagement \cite{pipattanasomporn2013load}. It is challenging to efficiently implement DR due to the highly varied energy load shapes across buildings. It is therefore important to learn and predict consumer behavior. Advanced metering infrastructure (AMI) and smart meters have been adopted to provide fine-grained (hourly or sub-hourly) energy data for advanced analytics to enhance the efficiency of power network operations. Electricity load shapes segmentation from smart meters, through clustering techniques, could inform decision-makers about different patterns of energy use towards improving resource allocation in power systems~\cite{wang2018review,rajabi2020comparative}.
With the wide adoption of smart meters nationwide~\cite{leiva2016smart}, and the availability of hourly and sub-hourly data, segmentation methods have been used to find similar patterns of electricity consumption behavior of customers. This task can be applied in different capacities to improve power system operation including DR programs implementation~\cite{kwac2014household}, load forecasting~\cite{quilumba2014using}, tariff determination~\cite{feng2019smart}, and renewable integration~\cite{afzalan2019semantic,xu2017household}. The primary objective of the customer segmentation task is to transform a large library of load shapes (i.e., all profiles in the database) into representative daily energy use routines, which altogether define the typical behaviors of the entire customer base.

A review of the literature shows that different efforts have mainly adopted clustering techniques and considered their output self-evident from an application standpoint, without further verification of how the resultant clusters properly reflect the actual energy behaviors. Specifically, averages of all profiles, associated with each cluster, have been treated as representations of a distinct energy use behavior, although the average representation may not properly represent the actual shapes of profiles in a cluster. This assumption could result in two problems: 1) the accuracy of the segmentation will be low and customers will be assigned to the clusters that do not reflect their energy behavior realistically, and 2) some of the representative energy use behaviors, which have considerable frequency in the database, might not be reflected in any of the clusters. As a result, certain characteristics that are important to the utilities and decision-makers in the process of load balancing and planning, including the distribution of relative peak demands in the load shapes and the magnitude of peaks, may not be reflected in the segmentation results.

Given these shortcomings, we introduced a segmentation approach that accounts for compatibility of consumption patterns across load shapes while capturing both temporal shapes and peak demands of usage profiles. We have proposed a two-stage clustering of daily electricity load shapes by initially overpopulating the set of clusters for preserving accuracy and further merging similar clusters by accounting for cluster shape alignment. Evaluation is done on real-world smart meter data using various metrics based on measuring error in clustering and correlation analysis. The key contributions of this work include (1) a proposed household electricity load shape segmentation for fine-grained representation of load profile patterns and peak demands, and (2) introducing quantified evaluation of segmentation quality, which improves on most prior efforts in power systems that considered the segmentation performance self-evident without investigating related assessment metrics.

 The rest of the paper is organized as follows: section~\ref{LR} presents the related work, section~\ref{ProblemStatement} defines the problem, section~\ref{method} discusses the proposed approach, section~\ref{results} presents the results, and sections~\ref{discussion} and~\ref{conclusion} conclude the paper.

\section{Related Work}
\label{LR}
The concept of load shapes segmentation has received increasing attention with the proliferation of smart meters. Initial efforts~\cite{chicco2006comparisons,chicco2003customer} have leveraged clustering techniques for classifying customers based on their load shape patterns. Different algorithms, including K-means \cite{kwac2014household}, Self-Organizing Map (SOM) \cite{verdu2006classification}, hierarchical clustering~\cite{chicco2012overview}, and Expectation-Maximization (EM)~\cite{coke2010random} have been investigated for the task of customer segmentation across tens to hundreds of households~\cite{kwac2014household,wijaya2014consumer}.
Several recent efforts in household electricity segmentation have focused on addressing specific applications. In a class of studies, the use of big data, and the efficiency of clustering at the city-scale have been investigated~\cite{iyengar2016analyzing, fontanini2018data}. Therefore, due to the increasing size of datasets, the applicability of dimensionality reduction techniques such as Principal Component Analysis (PCA) and Symbolic Aggregate Approximation (SAX)~\cite{rajabi2020comparative}, in addition to feature extraction from the attributes of interests from load shapes (such as peak distributions or key timeframes) have been investigated~\cite{haben2015analysis,afzalan2019semantic}. Furthermore, to mitigate the impact of noisy and unequal time-series with incomplete information in the real-world scenarios, the model-based approach, which accounts for the phase shift and time lag has been proposed~\cite{motlagh2019clustering}. In some recent attempts, investigation of segmentation on disaggregated (i.e., individual load level) data, including Air Conditioning (AC) has been carried out for DR applications~\cite{malik2019appliance,lin2017clustering}.

As a core assumption in household electricity segmentation and its potential applications, the clustering output (both in terms of cluster quality and the number of clusters) has been assumed to reasonably represent the energy behavior patterns of the entire customer base. Such an assumption has been deemed either as self-evident without further investigation \cite{rhodes2014clustering}, investigated through common cluster validation indices (CVI) \cite{mcloughlin2015clustering,rajabi2020comparative,chicco2012overview}, or justified through empirical inspection~\cite{iyengar2016analyzing}. The application of common CVIs has been deemed to be viable for quality assessment of the clustering outcome at the first sight. Examples of typical CVIs include Bayesian Information Criterion (BIC) \cite{bic}, Silhouette index \cite{sil}, and Davies–Bouldin index (DBI) \cite{db}. However, as outlined in \cite{iyengar2016analyzing}, such generic statistical indicators for model selection will not necessarily work for the electricity segmentation task. In other words, proper metrics that indeed measure the compatibility of individual load shapes with their representative clusters with respect to their associated profiles have been mainly ignored in the evaluation processes~\cite{xu2017household}.

As the trend in the literature shows, the household electricity segmentation efforts have mainly overlooked the importance of recovering nuances of load shapes with distinct temporal patterns. Therefore, in this work, we focused on the fine-grained load shape cluster representation with distinct temporal patterns and peak demand magnitude through two-stage segmentation. 
\section{Preliminaries and Motivation}
\label{ProblemStatement}
Due to the highly varied patterns of energy consumption across different households on a daily basis, the task of segmentation can become challenging. In this section, the problem statement is presented through a quantitative case-study. We first demonstrated how common cluster validation indices (CVIs) for segmentation can lead to coarse-level representation of load shapes. We further highlighted the importance of recovering distinct load shapes from the clustering process.
\subsection{Dataset description}
\begin{figure}[]
\centering
\includegraphics[width=\linewidth]{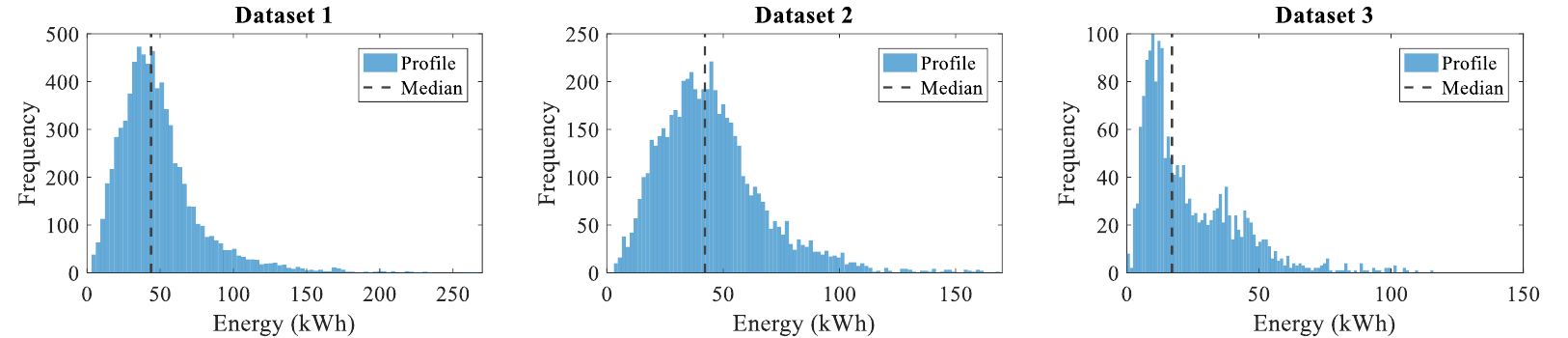}
\caption{\label{data}Distribution of energy consumption among all profiles.}
\end{figure}

In this study, we used the data from the Pecan Street Project [41], which is an ongoing campaign in energy-efficiency initiative through equipping residential buildings with metering devices. Here, we used a subset of a dataset that was collected from residential buildings in Austin, TX and Boulder, CO during July and August 2015. Table \ref{dataTable} presents the characteristics of the datasets. The resolution of data was one sample per 15-minute. Therefore, each daily profile contains 96 data points. The duration of data collection for each dataset was 60 days. Three datasets were considered to impose adequate variations in the energy consumption styles. However, since the households were primarily located in Austin, TX, we divided the households in that geographical locations into two datasets.  
\begin{table}
\centering
\caption{Characteristics of the dataset}
\label{dataTable}
\resizebox{\columnwidth}{!}{
\begin{tabular}{|c|c|c|c|c|c|}
\hline
Dataset & Location & Number of households & \# of daily profiles \\ \hline
Dataset 1 & Austin, TX & 129 & 7535 \\ \hline
Dataset 2 & Austin, TX & 100 & 5676 \\ \hline
Dataset 3 & Boulder, CO & 31 & 1790 \\ \hline
\end{tabular}
}
\end{table}
Figure \ref{data} shows the distribution of total daily energy consumption for all observations in different datasets. The median energy consumption for dataset 1, dataset 2, and dataset 3 is 44 kWh, 42kWh, and 17kWh, respectively. 
\subsection{Problem statement description}
\begin{figure*}[!t]
\centering
\includegraphics[width=0.9\linewidth]{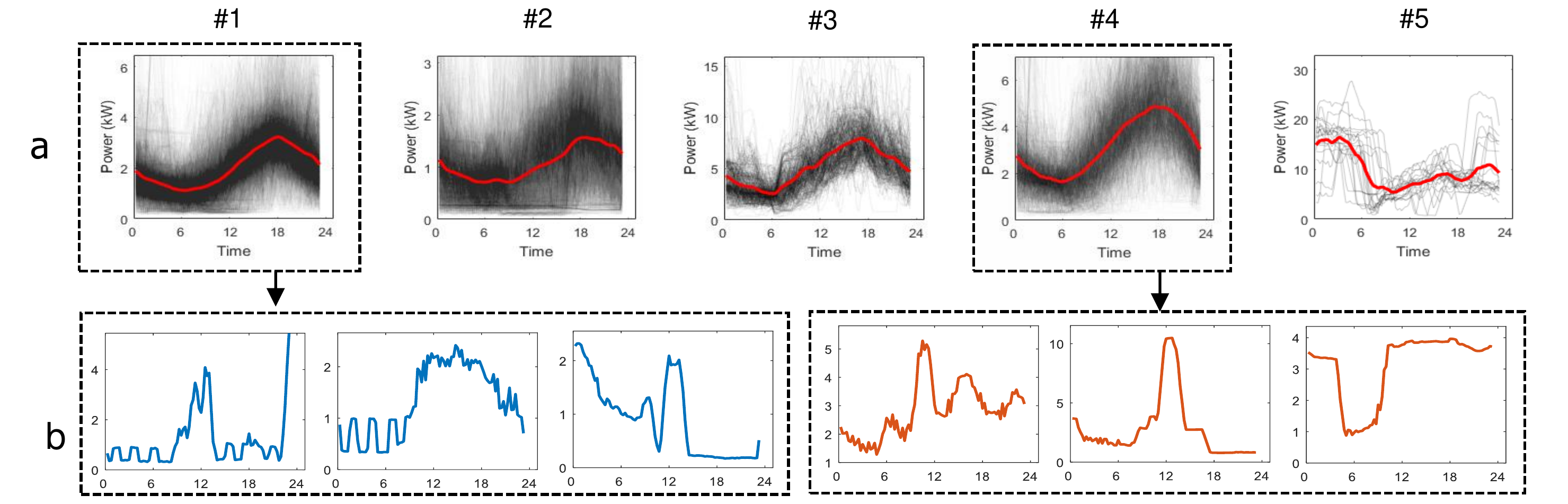}
\caption{\label{segExample}Segmentation of $\sim$8000 load shapes using K-means: a) five clusters representing the entire dataset, b) examples of individual daily profiles for clusters 1 and 4 that are misclassified.}
\end{figure*}

To demonstrate the problem in a generic form, a common clustering outcome on daily profiles has been presented. Figure \ref{segExample} shows the clustering results of $\sim$8000 daily load shapes clustered into 5 groups using K-means algorithm, which is commonly applied in the literature for the datasets of similar size and nature. In Figure \ref{segExample}(a), each subplot shows the temporal shape of each cluster, and the vertical axis is the power magnitude (in kW). The red line is the centroid of each cluster, which is obtained by averaging all daily profiles (in the order of hundreds or thousands) associated with the cluster. In Figure \ref{segExample}b, several examples of daily profiles associated with cluster 1 and 4 are shown. In the literature, the segmentation results were typically presented similar to what shown in Figure \ref{segExample}a, without further investigation of cluster quality. Therefore, the validity of clustering results were assumed as self-evident, and no further investigations were carried out to show how representative each cluster is with respect to its associated members. However, as shown in Figure~\ref{segExample}b, examples of load shapes for \#1 are retrieved that do not resemble the shape of its centroid. Similarly, a number of load shapes in cluster \#4 are presented (Figure \ref{segExample}b) that do not resemble their cluster centroid (Figure \ref{segExample}a). Considering that these examples of load shapes in Figure \ref{segExample}(b) have considerable density in the entire dataset, it is important to have their own clusters to represent distinct energy behaviors. While one solution is to increase the number of clusters to allow for better representation of load shapes, this could come at the cost of obtaining a large number of clusters with high similarity, that contradicts the objective of segmentation.
Based on this discussion, it could be seen that variations in temporal patterns of load shapes and their peak demands make the segmentation task challenging. Specifically, this problem becomes more important as the scope of segmentation gets larger both with higher number of households and historical days. Therefore, it is important to perform efficient segmentation on such datasets to allow for the trade-off between representing typical energy consumption patterns and maintaining interpretable number of clusters.
\subsection{Investigation of typical CVIs}
\label{CVI}
Segmentation is commonly done using a small number of clusters. This is intuitive to ensure the ease of interpretation as a primary criterion for segmentation. To justify the number of clusters, cluster validation indices (CVIs) are commonly used. Here, to show the challenge of using typical CVIs and show the complexity in clustering patterns, we used several well-known CVIs, namely, the Davies-Bouldin Index (DBI) \cite{db}, Silhouette (SIL) \cite{sil}, Calinski Harabasz (CH) \cite{ch}, and Within Cluster Sum of Squared (WCSS) error \cite{sse}. CVIs definitions are provided in Table \ref{tab:tableCVI}.
\begin{table*}[!t]
\centering
\caption{CVI description.}
\label{tab:tableCVI}
\resizebox{\textwidth}{!}{
\begin{tabular}{|l|l|l|l|}
\hline
CVI & Definition & Selection criterion & Ref \\ \hline
Davies Bouldin Index (DBI) & $\frac{1}{K}\sum_{i}max_{j,j \neq i}\{\frac{\frac{1}{\parallel C_i \parallel}\sum_{x \in C_i}d(x,\mu_i)+\frac{1}{\parallel C_j \parallel}\sum_{x \in C_j}d(x,\mu_j)}{d(\mu_i,\mu_j)}\}$ & Minimum & \cite{db} \\ \hline
Silhouette (SIL) & $\frac{1}{K}\sum_{i}\{ \frac{1}{\parallel C_i \parallel} \sum_{x \in C_i} \frac{b(i)-a(i)}{max(b(i),a(i))} \}$ & Maximum & \cite{sil} \\ \hline
Calinski-Harabasz Index (CHI) & $\frac{\sum_{i} \parallel C_i \parallel *d^{2}(\mu_i,\mu_j) / (K-1)}{\sum_{i} \sum_{x \in C_i} d^{2}(x,\mu_j) / (N-K)}$ & Maximum & \cite{ch} \\ \hline
\multicolumn{1}{|l|}{Within Cluster Sum of Squared (WCSS)} & $\sum_{i} \sum_{x \in C_i} \parallel x-\mu_i \parallel^{2}$ & Elbow & \multicolumn{1}{l|}{\cite{sse}} \\ \hline
\end{tabular}
}
\begin{flushleft}
Notation: $K$: number of clusters; $C_i$: Cluster $i$; $\parallel C_i \parallel$: Number of load shapes in $C_i$; $\mu_i$: centroid of $C_i$; $N$: total number of load shapes; $a(x)=\frac{1}{\parallel C_i\parallel -1}\sum_{j \in C_i, i \neq j}d(i,j)$, $b(x)=min_{k \neq i} \frac{1}{\parallel C_i \parallel} \sum_{j \in C_k}d(i,j).$
\end{flushleft}
\end{table*}
We measured the CVIs for four common clustering methods (Self-Organizing Map (SOM), K-means, Fuzzy c-means,and hierarchical clustering). To choose the number of clusters ($K$), CVI is measured over a range of $K$ values, and a selection criterion (Table \ref{tab:tableCVI}) is used on CVI values.
We compared the range of $K$={5,10,15,20,…,120} to have a reasonable estimation of low to high number of clusters. We used 1 cluster increments in the commonly used range 5 to 10 (i.e., {5,6,7,8,9,10}), and for more than 10, we used 5 cluster increments till $K$(end)=120 (i.e., {10,15,20,…,120}). 
For each value of $K$, we cluster and measure CVIs.
Figure \ref{cvi} presents the CVIs for Dataset 1, with one subplot representing one CVI. The trend of increase/decrease for all 4 CVIs (each subplot) is almost consistent across datasets. Therefore, we presented the interpretation for Dataset 1 below, and the findings were consistent with other datasets as well.
DBI metric uses a min-rule for selecting the proper $K$. As the results show, for all methods, a low value of $K$=5 leads to low DBI values. It must be noted that DBI values for fuzzy c-means showed to be very high. Therefore, we did not present its values in the first subplot to avoid masking the trend of changes for other methods. SIL metric uses a max-rule for selecting the $K$. Similar to the previous case, the lowest value of $K$=5 is estimated. CHI metric uses a max-rule for selecting the $K$. We select $K$=5 for this metric. However, for WCSS, that follows the elbow criterion for selecting the $K$, $K$=30 is a reasonable value for hierarchical, SOM, and K-means.

When it comes to algorithms, for DBI and SIL metrics, K-means shows a marginal improvement over SOM and hierarchical clustering. For CH metric, SOM and K-means show an equally better performance. For the WCSS metric, hierarchical clustering has the lowest values and seems to have a better performance. Therefore, there is no consistency in selecting a superior
approach by considering combinations of CVIs and clustering
algorithms. Nonetheless, to visually evaluate the performance of common algorithms based on optimum CVIs, an assessment of the clustering results is presented in the next section. Table \ref{tableClus} shows optimum $K$ values, inferred based on CVIs for each dataset investigated in the next section.
\begin{figure}
\centering
\includegraphics[width=1\linewidth]{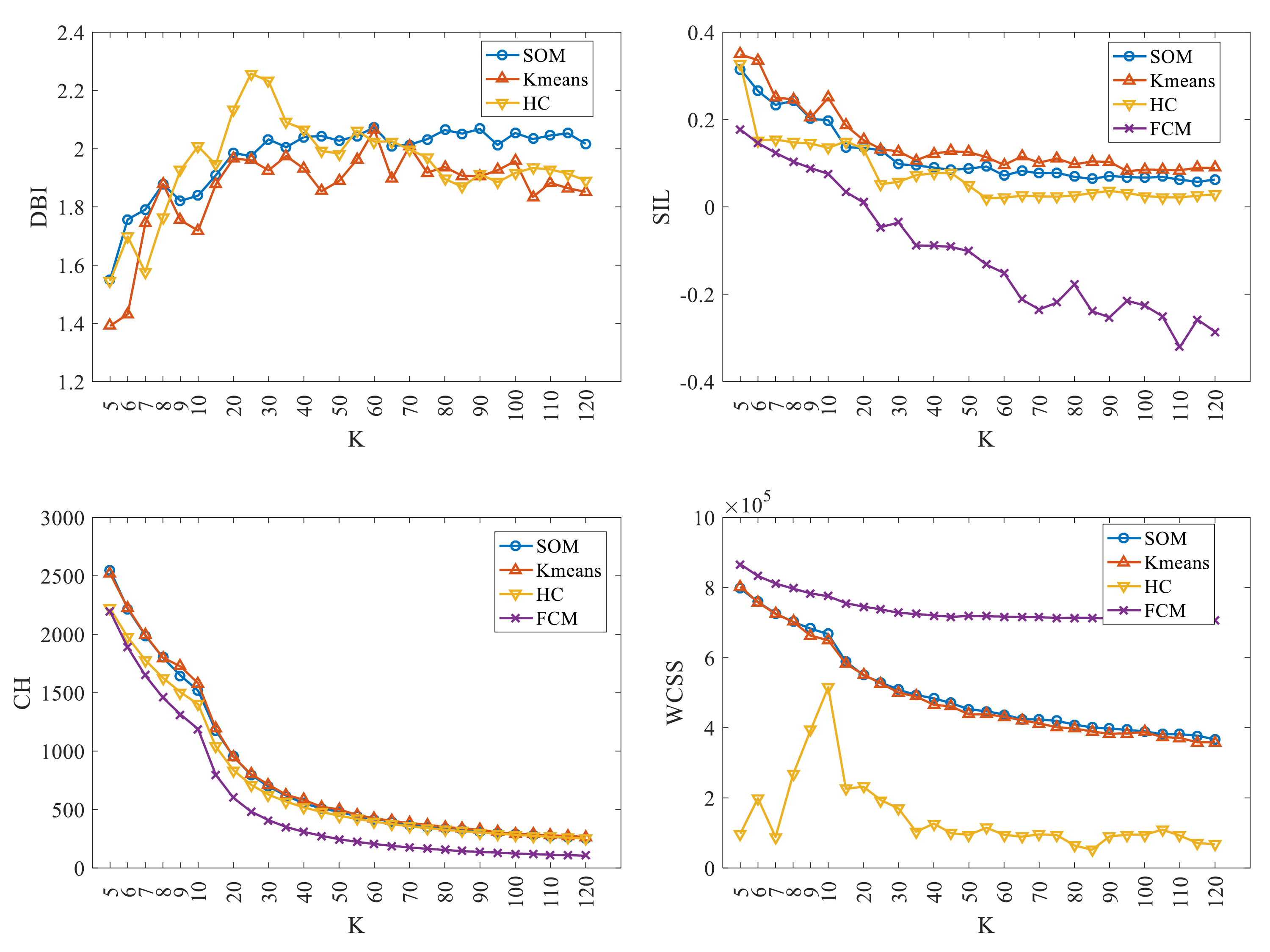}
\caption{\label{cviFIG} CVI for different clustering algorithms.}
\end{figure}
\begin{table}[]
\caption{K values for different datasets and algorithms inferred by using the optimum CVIs.}
\label{tableClus}
\begin{tabular}{|c|c|c|c|}
\hline
\textbf{Algorithm} & \textbf{K (Dataset 1)} & \textbf{K (Dataset 2)} & \textbf{K (Dataset 3)} \\ \hline
K-means & 5 & 8 & 6 \\ \hline
HC, SOM & 30 & 25 & 25 \\ \hline
\end{tabular}%
\end{table}

\subsection{Comparison of clustering results}
\label{compare}
\begin{figure}
\centering
\includegraphics[width=1\linewidth]{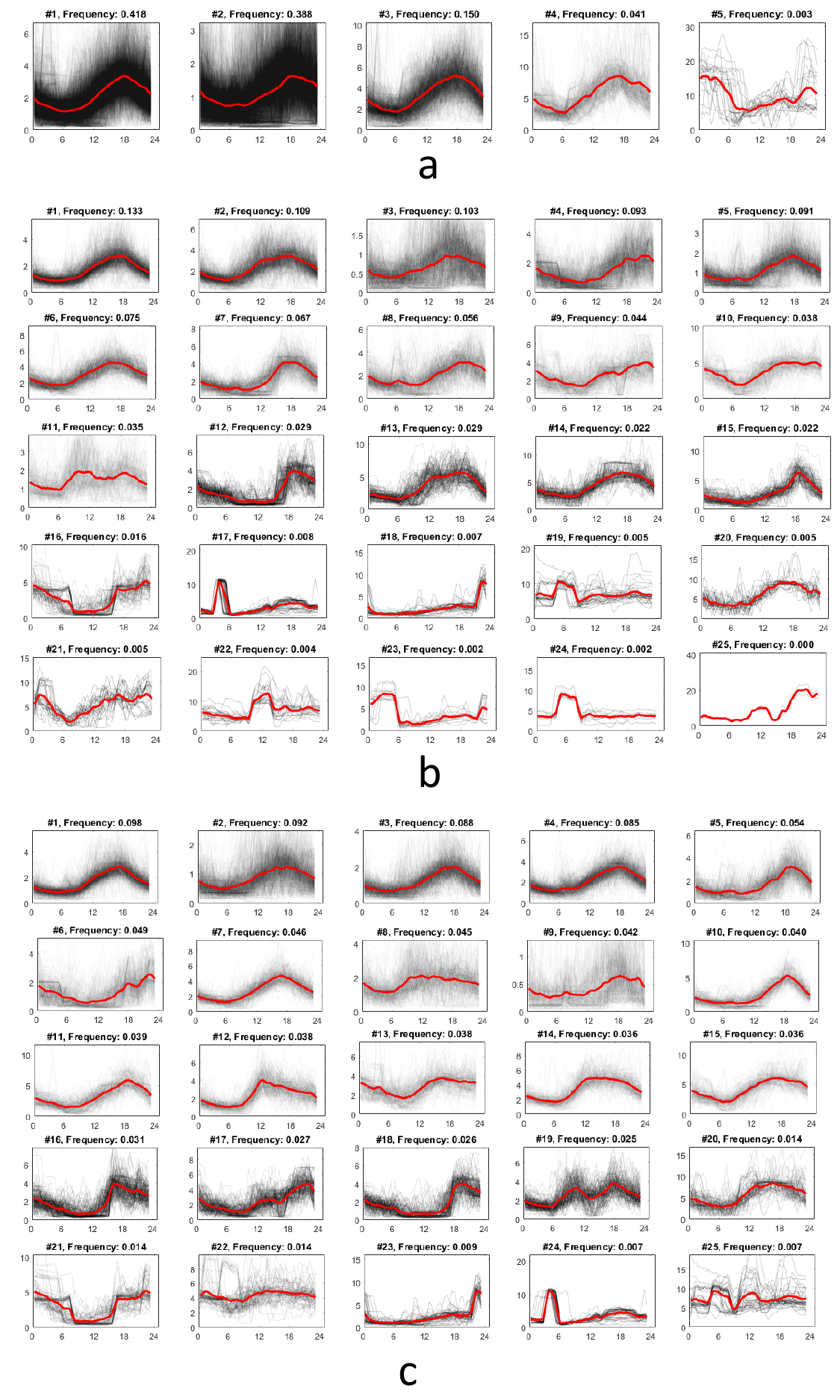}
\caption{\label{cvi} Cluster comparison for dataset 1: a) K-means, $K$=10, b) HC, $K$=30, c) SOM, $K$=30. The vertical axis is ‘Power (kW)’ and the horizontal axis is ‘Time (hr)’.}
\label{clusCompare}
\end{figure}

Form cases in Table \ref{tableClus}, Figure \ref{clusCompare} presents the clusters for dataset 1 as an example demonstration. Each subplot is one cluster, and the red curve is the cluster centroid, averaged over all profiles. The frequency value above each subplot shows the occurrence rate of the cluster within the entire dataset. In Figure \ref{clusCompare}(a), with $K$=5, four out of five clusters have peak demand around 18:00, but with different peak magnitudes. Cluster 5 seems to be an outlier, due to its considerable high peak usage in addition to its very low frequency (less than 1\% of the data). In Figure \ref{clusCompare}(b) and Figure \ref{clusCompare}(c), with a higher population of clusters, more distinct energy patterns were revealed, while their presence has been masked by selecting a lower number in Figure \ref{clusCompare}(a). Examples include clusters 4, 11, and 17 in Figure \ref{clusCompare}(b), and clusters 8, 19, and 21 in Figure \ref{clusCompare}(c). Furthermore, a comparison between hierarchical clustering and SOM with equal number of clusters (Figure \ref{clusCompare}(b) and Figure \ref{clusCompare}(c)) show that hierarchical clustering emphasizes on identifying clusters with very low frequency (4 clusters with frequency of less than 0.5\%) while SOM forms clusters with higher frequency. As shown in Figure \ref{clusCompare}, although CVI selects $K$=5, using such a low value will result in a very coarse-level representation of load shape groups. Specifically, the cluster centroids, which altogether are supposed to encompass the energy use behavior of the community, might not reflect the temporal shape/peak demands of their associated members. 

Given the variations in the nature of CVIs, using different metrics results in selecting different $K$ values for different algorithms. Generic CVIs might not work well for this domain-specific problem because several groups of energy patterns with distinct loads may not be recovered without selecting a large number of clusters, while their shapes/magnitude characteristics could be of interest to energy planners/utilities. Therefore, the next section introduces an approach to tackle this problem.
\section{Two-stage clustering on household electricity load shapes}
\label{method}
\begin{figure}
\centering
\includegraphics[width=0.4\textwidth]{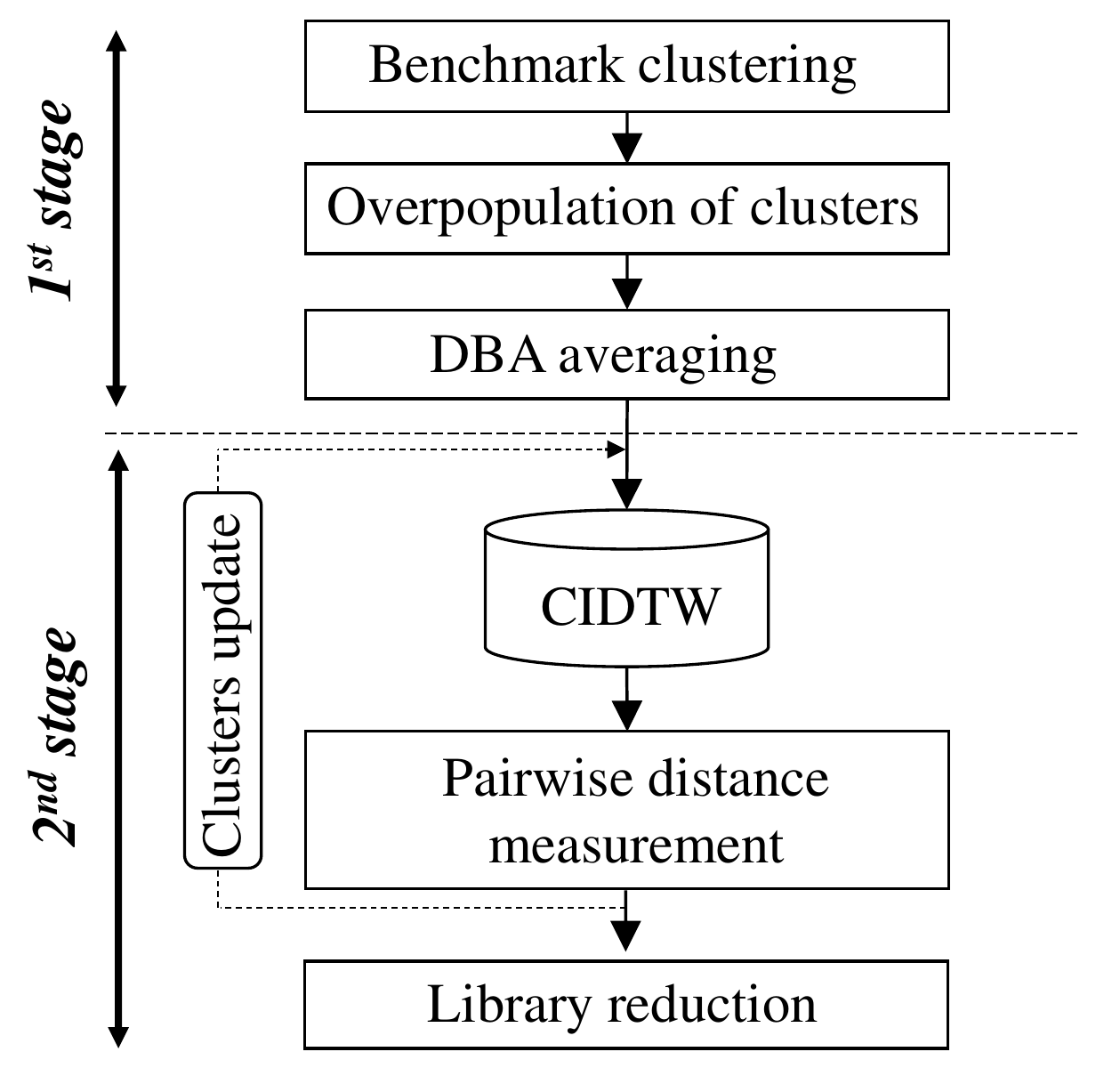}
\caption{\label{twoStage}Two-stage clustering framework. Benchmark clustering is an arbitrary clustering technique (e.g., k-means). Acronyms: DBA (Dynamic Time Warping Barycenter Averaging); CIDTW (Complexity-Invariant Dynamic Time Warping).}
\end{figure}

As discussed in Section~\ref{compare}, optimizing the number of clusters on CVIs could result in an unreasonably low number of clusters. On the other hand, selecting a high number of clusters could improve the representation of various load shapes. However, it could affect the interpretation of the outcome due to presence of multiple correlated clusters. To account for this trade-off, a two-stage clustering approach is introduced that initially overpopulates clusters to improve the accuracy of load shape representations and further merge the closely related ones to improve the interpretability.
The general framework is shown in Figure~\ref{twoStage}. The objective is to reduce the initial cluster library with fine-grained representation without losing the essential information in load shape patterns/peak demands. In the first stage, a large number of clusters is generated, which are then transformed with a time-series averaging technique to represent cluster centroids. In the second stage, pairwise distances of clusters are calculated with a distance measure that accounts for shape alignment between cluster centroids to merge the similar ones. Each component is described next.
\subsection{First stage: Initial cluster representation}
\subsubsection{Overpopulation of clusters} 
To provide a diverse set of load shapes that captures a wide selection of the possible profiles, a representative of the true distribution of the profiles in the entire dataset is needed. In the first stage, an initial value of $K^'$ for the number of clusters is considered ($K^'>K$). The selection of $K^'$ can be performed by measuring the WCSS error, such that increasing $K^'$ do not cause a considerable change in the SSE. To this end, the elbow curves for WCSS (described in section~\ref{CVI}) is used for the selection of $K^'$. Using any clustering technique, $K^'$ groups are populated.

\subsubsection{Cluster representation}
Since each cluster may include thousands of observations, a proper representation for each cluster, that resembles the content for each group is required. The most intuitive way for cluster representation is the Euclidean averaging (simple averaging of each sample across all observations). However, Euclidean averaging may result in a centroid which is dissimilar to any of its associated time-series in a given cluster~\cite{petitjean2014dynamic}. To this end, we employed the Dynamic Time Warping Barycenter Averaging (DBA) technique~\cite{petitjean2011global}. DBA is a time-series averaging technique that preserves the nuances of variations in individual profiles. In contrast to conventional time-series averaging which may result in centroids that differ from original time-series, DBA uses an expectation maximization approach by refining the medoid of each group through finding the best set of alignments within each group through iterations. In each iteration~\cite{petitjean2011global}: \newline 1) DTW distance between each profile ($ x \in C_i$) and the temporary average centroid ($\mu_i^'$) is measured. This process is updated in each iteration to find the association of each sample ($t \in {1,2,...,T}$) of the centroid ($\mu_i$) with respect to samples ($t \in {1,2,...,T}$) of individual profiles associated with cluster $i$.
2) Updating each sample of the centroid as the barycenter of samples associated with it from the previous step. 
The barycenter function is defined as:
\begin{equation}
barycenter(\xi_1,\xi_2,…,\xi_m)=(\xi_1+\xi_2+…+\xi_m)/m 
\end{equation}
Using previous iteration average centroid ($\mu_i^'$), the \textit{t}-th sample of the current iteration average centroid ($\mu_i$) is defined as:
\begin{equation}
   \mu_i (t)=barycenter(assoc(\mu_i^' (t)))   
\end{equation}
Here, $assoc$ function associates each sample of the $\mu_i$ to the samples (one or more) of the profiles based on calculating a Dynamic Time Warping (DTW) distance.

\begin{figure}
\centering
\includegraphics[width=1\linewidth]{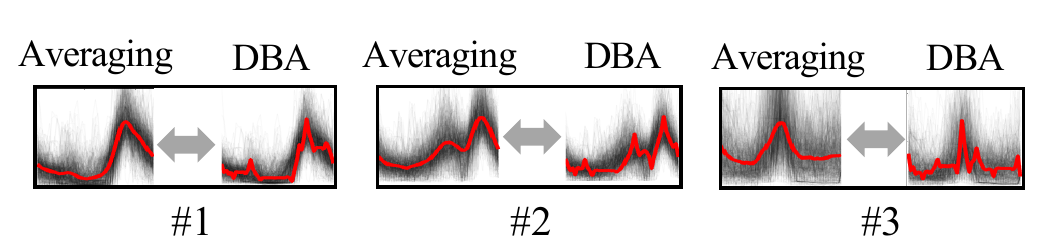}
\caption{\label{DBA} DBA averaging versus conventional averaging.}
\end{figure}

Considering $K^'$ initial clusters, DBA is applied to the content of each one to obtain the centroids $\mu_i$.
To show the impact of DBA versus conventional averaging, we presented a set of clusters in Figure~\ref{DBA}. As can be seen, DBA could enhance the representation of centroids in each cluster by sharpening the peaks and valleys reflected in profiles.
\subsection{Second stage: Cluster merging}
\begin{figure}
\centering
\includegraphics[width=.6\linewidth]{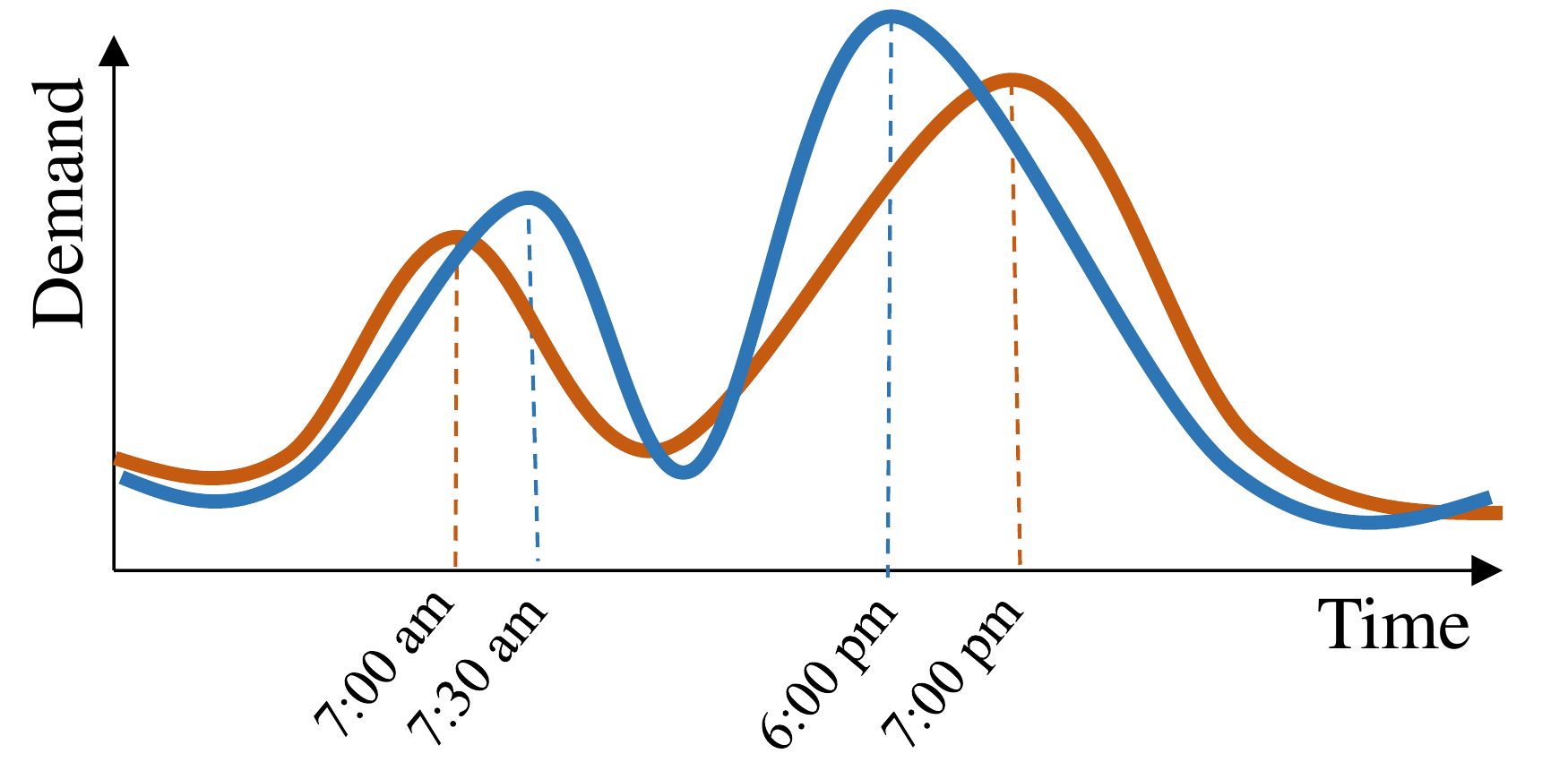}
\caption{\label{DTW} Load shapes with similar behaviors and temporal shift.}
\end{figure}

In this stage, the clusters that are highly similar in shape and magnitude are merged to reduce the final library size. Given the initial cluster size of $K^'$ in the first stage, the objective is to reduce the library size to $K$. We used an iterative merging process to transform the dataset from $K^'$ to $K$ clusters ($K$ could be interpreted from WCSS elbow curve). In each iteration, the matrix of similarity measure between cluster $i$ and $j$ is constructed and the pairs that are closest are merged ($K^'\to K^'-1$). The process is continued till $K^' \to K$.
Due to the nature of electricity demand profile datasets, it is important to employ a robust measure for merging the time-series that accounts for the inherent small time shift in similar load shapes. Figure~\ref{DTW} shows an example of two household load shapes that have similar energy use patterns (double demand peaks in the morning and evening) but are relatively different in the time of peak demand ($\sim$1 hour difference in peak timing). Typical similarity metrics such as Euclidean distance fail to capture similarity in such cases. We employed the Complexity-Invariant Dynamic Time Warping (CI-DTW) as the distance measure~\cite{cidtw}. CI-DTW is a DTW-based distance measure that is invariant to the complexity of time-series (e.g., number of peak and valleys). Therefore, it avoids matching pairs of simple objects that are subjectively apart from those with more complex patterns with similar shapes~\cite{cidtw}.
\subsubsection{CI-DTW}
CI-DTW \cite{cidtw} is a variation of DTW distance measure. In DTW \cite{dtw}, the optimal alignment of the two time-series ($P=\{p_1,p_2,…,p_T\}$ and $Q=\{q_1,q_2,…,q_T\}$) is recursively found by calculating the cost defined by:
\begin{equation}
    D(i,j)=\delta(p_i,q_j )+min\begin{cases}
                 D(i-1,j-1)\\
                 D(i,j-1)\\
                 (D(i-1,j)\\
              \end{cases}
\end{equation}
in which $\delta(p_i,q_j)$ is the distance between the $i-th$ sample of $P$ and the $j-th$ sample of $Q$. The above equation can be calculated by dynamic programming. Upon measuring the DTW distance, CI-DTW is defined as:
\begin{equation}
CIDTW = DTW(P,Q)*CF(P,Q)                             
\end{equation}
in which $CF(P,Q)$ is a correction factor as follows:
\begin{equation}
  CF(P,Q)=\frac{max(CE(P),CE(Q))}{min(CE(P),CE(Q))}  
\end{equation}
and $CE(A)$ is a complexity estimate as:
\begin{equation}
CE(A)=\sqrt{\sum_{i=1}^{t-1}(a_i-a_{i+1})^2}         
\end{equation}
\subsubsection{Iterative merging}
\begin{figure}
\centering
\includegraphics[width=\linewidth]{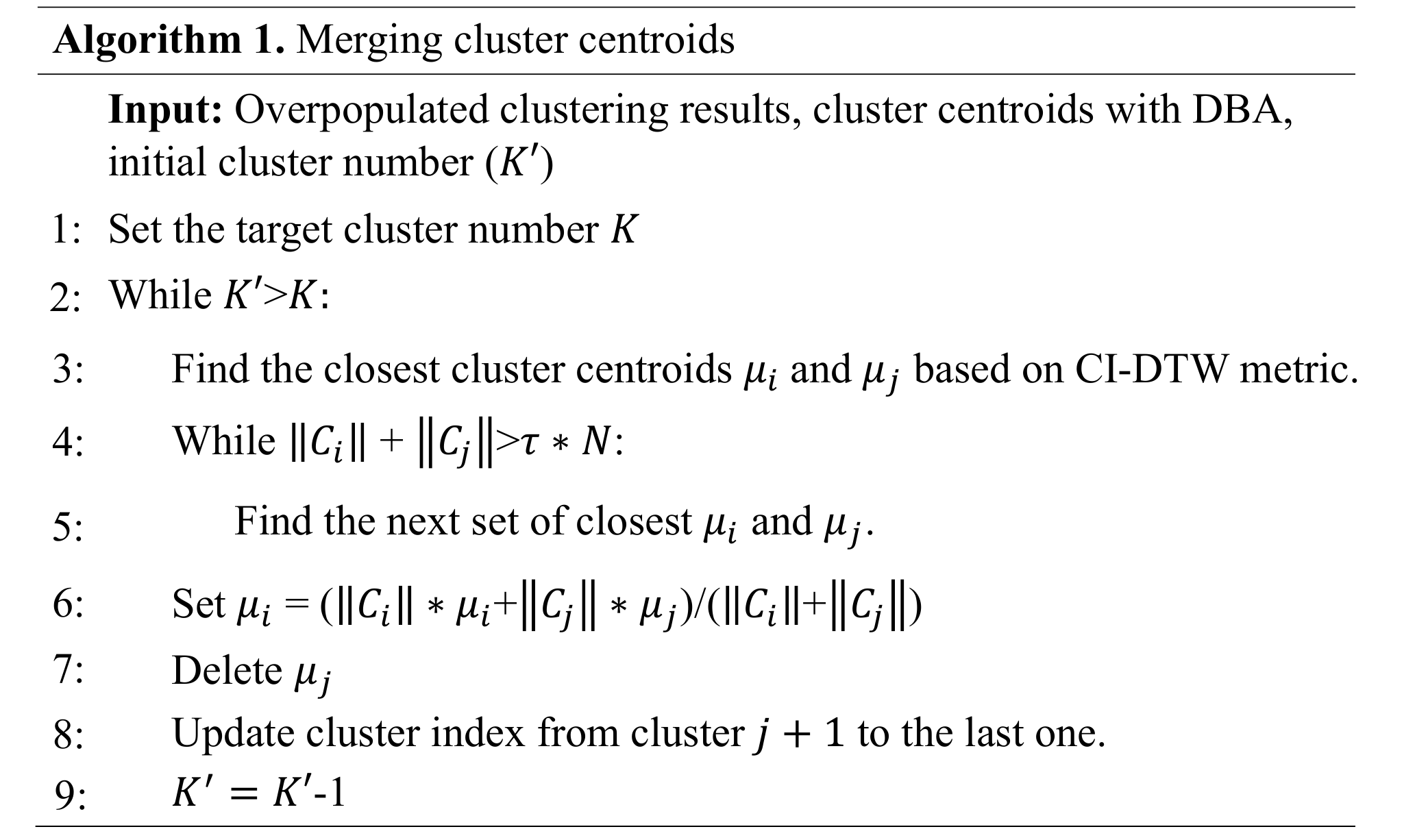}
\caption{\label{pseducode}Pseudocode for cluster merging.}
\end{figure}

Using the CI-DTW distance measurement between cluster centroids, we merge the closest cluster pair ($i,j$) in each iteration, and update the content and centroid of the remaining ones. A control parameter ($\tau$) is considered as the maximum cluster density upon merging $i,j$ such that:
\begin{equation}\parallel C_i \parallel + \parallel C_j \parallel \leq \tau*N\end{equation}
Where $\parallel C_i \parallel$ and $\parallel C_j \parallel$ are the \# of profiles associated with clusters $i$ and $j$, respectively, and $N$ is the total \# of profiles in the dataset. Therefore, $\tau$ controls post-merging cluster size to avoid the formation of highly dense clusters. In the presented results, $\tau$=0.2 (i.e., 20\% of the dataset) was considered. Figure \ref{pseducode} shows the pseudocode for the second stage cluster merging. 
\section{Results and discussion}
\label{results}
\subsection{Visualization and empirical investigation}
\begin{figure}
\centering
\includegraphics[width=\linewidth]{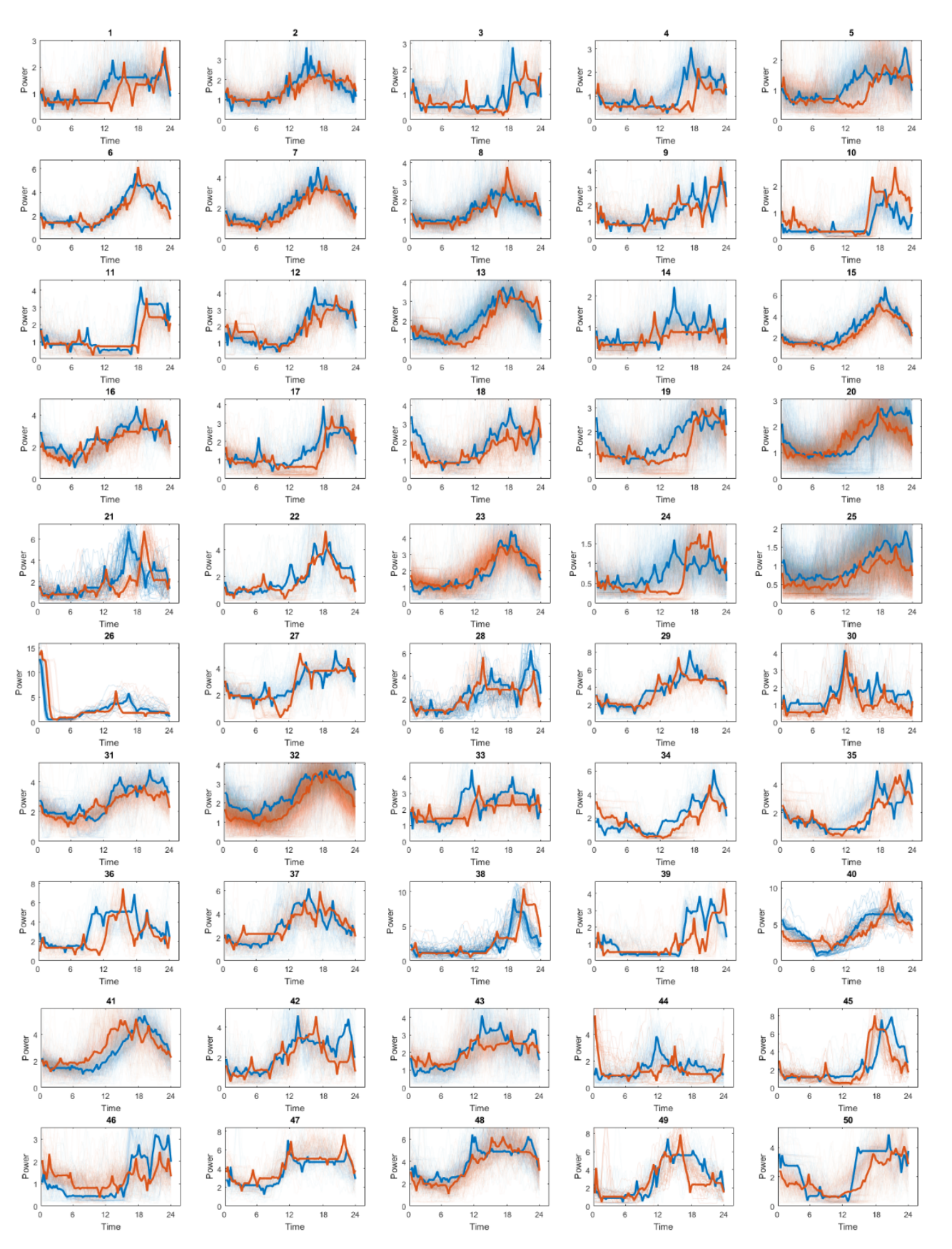}
\caption{\label{merge} Merged cluster pairs at different iterations (Initial \#clusters=90; Final \#clusters (stopping criterion)=40; \#iterations=50). Iteration \# above subplots.}
\end{figure}

\begin{figure}
\centering
\includegraphics[width=\linewidth]{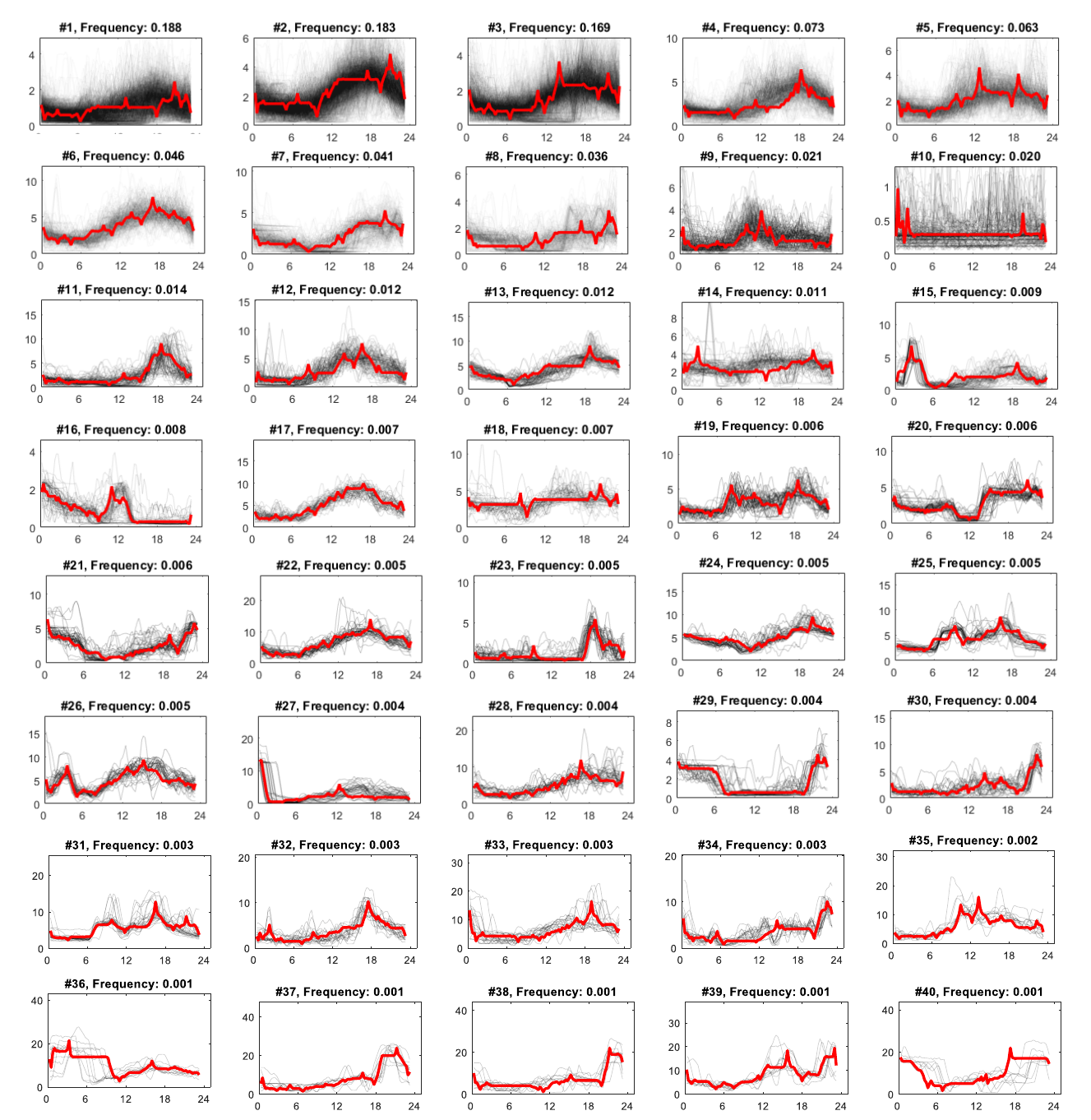}
\caption{\label{finalMerge}Two-stage clustering ($K^'$=90,$K$=40). Subplots' horizontal axis: time of day (hr); vertical axis: power (kW)}
\end{figure}

We apply the two-stage clustering approach described in section~\ref{method} using SOM and K-means as the first stage baseline since they outperform fuzzy c-means and hierarchical clustering, which tend to generate outliers. To estimate the initial number of clusters ($K^'$), the elbow curve for WCSS in Figure~\ref{cviFIG} was used. A set of $K^'$=\{50,70,90\}, spanning the range in which the error decline between subsequent $K$s was low, was considered for overpopulation of clusters. For the second stage, the final library size of $K$=\{10,20,30,40\} was used to study the impact of cluster merging at different levels.
Figure~\ref{merge} shows the pairs of merged clusters at different iterations ($K^'$=90; $K$=40; $K^'-K$=50). In this figure, SOM was applied for creating the initial cluster library. The value above each subplot is the iteration number. The results show that the selected clusters are subjectively close in both temporal shape patterns and peak magnitudes.
Figure~\ref{finalMerge} shows the final clusters for this example after merging. Empirical evaluation shows that clusters are well-separated and distinct in their temporal shapes and power magnitude. Furthermore, they accentuate the useful features in load shapes such as peak magnitude, peak timing, peak duration, and energy consumption (area under load shape curve), which could be of interest to utilities for planning customized energy programs. 
\subsection{Quantified investigation}
\textbf{Metrics.} We used the weighted average correlation ($WAC$) coefficient and $WCSS$ as the two quantified metrics for comparison. Both of these values reflect the compactness of clusters. Therefore, they resemble the extent to which the cluster centroids represent their associated profiles. $WAC$ first measures the correlation coefficient of each cluster centroid with respect to its associated profiles and uses the average value as the correlation indicator of each cluster. Thereafter, the frequency of each cluster is utilized to have the weighted average as a single correlation score. More specifically:
\begin{equation}WAC=\frac{\sum_{i=1}^{K}\parallel C_i \parallel * corr_i}{N}, WAC \in [0,1]\end{equation}
In which $WAC$ is the weighted average correlation and $corr_i$ is the average correlation coefficient for cluster $i$ as follows:
\begin{equation} corr_i=\frac{\sum_{x \in C_i} corr(x,\mu_i)}{\parallel C_i \parallel}\end{equation}
$WCSS$ error has also been defined in section~\ref{CVI}. A lower value of $WCSS$ indicates the higher compactness.
	
\begin{figure}
\centering
\includegraphics[width=\linewidth]{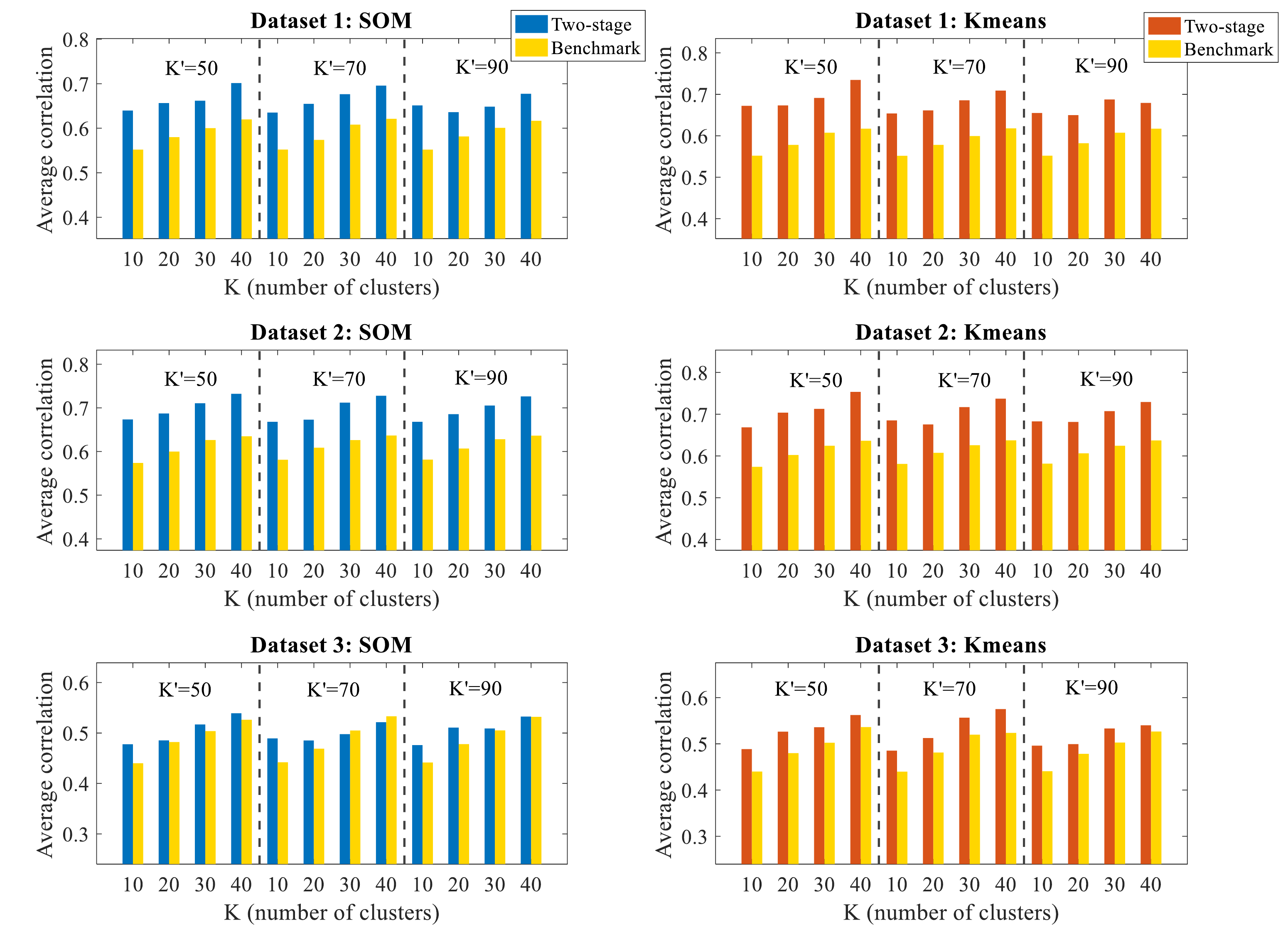}
\caption{\label{corr}Correlation coefficient comparison to benchmark. Left subplots: SOM; right subplots: K-means. Higher is better.}
\end{figure}

\begin{figure}
\centering
\includegraphics[width=\linewidth]{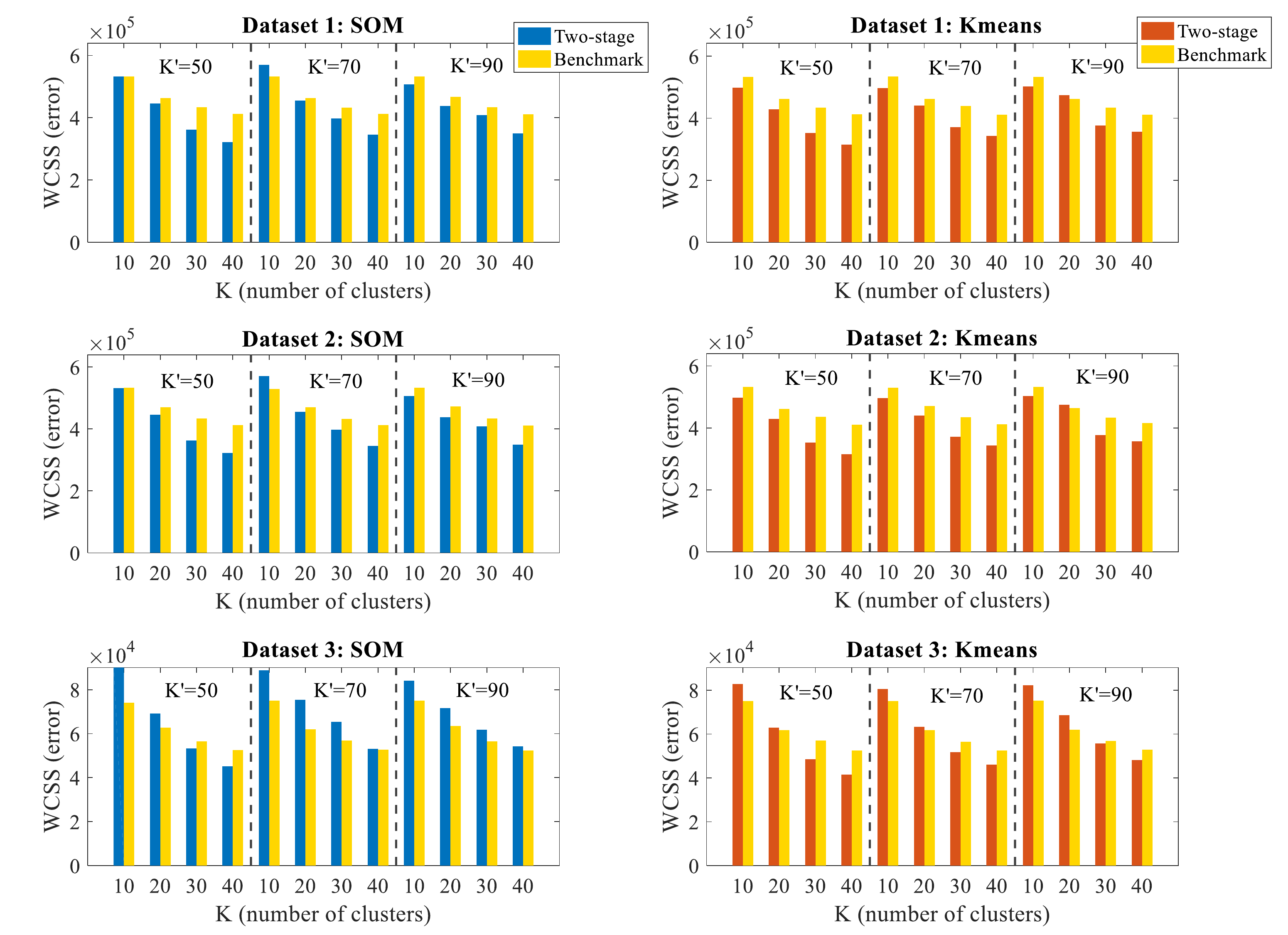}
\caption{\label{wcss} WCSS comparison to benchmark. Left subplots: SOM; right subplots: K-means. Lower is better.}
\end{figure}

Figure \ref{corr} presents the $WAC$ for different scenarios. Each row represents one dataset, and each column is one clustering method (SOM on left and K-means on the right). The two-stage bars show the final results after overpopulating the clusters (with $K^'$ values shown in the subplots) and then merging them up to $K$ clusters (shown on the horizontal axis). The benchmark bar represents the conventional clustering by directly selecting $K$ clusters. As the results show, using the two-stage approach improves the correlation in most cases. Specifically, it improves the average correlation by 8.2\%, 8.9\%, and 2.6\% for dataset 1, dataset 2, and dataset 3, respectively.
Figure \ref{wcss} shows the WCSS metric results. The two-stage approach results in lower error values in most cases, indicating the higher compactness of clusters. Specifically, it reduces the WCSSE average by 9.3\%, 9.5\%, for datasets 1 and 2, respectively. However, for dataset 3, an average of 3.4\% increase was observed. A possible interpretation for the error increase is lower size of dataset 3 (Table \ref{dataTable}). Therefore, the initial $K'$ values used in Figure \ref{cviFIG} may not be appropriate for this dataset. Other solutions~\cite{al2020novel} for selecting the appropriate cluster number might address this issue.
\section{Discussion for customized energy programs}
\label{discussion}
Based on the findings in section \ref{results}, the proposed two-stage clustering could help accentuate useful load shape features such as peak magnitude, peak timing, peak duration, and energy consumption, which are important to electricity providers. To provide some context on the implication of findings, we interpreted results of Figure \ref{finalMerge} from the application standpoint.
Demand Response (DR) events are typically scheduled for the evening timeframe, in which the net demand of the network is excessively high. Based on Figure \ref{finalMerge}, clusters 4, 5, 6, 7, 11, 13, 17, 19, 22, 33 have sharp peaks during 5pm-7pm timeframe, which makes them potential candidates for DR events during those timeframes. The implementation of DR could be made by partial load shedding or load shifting. Clusters 2, 8, 21, 30, 34, and 37 have also a sharp peak at a later timeframe, which makes them potential candidates for DR for timeframes after 8pm. Especially, clusters 30, 32, 33, 34, and 37 have considerably high usage (peak demand of more than 10kW), whose peak consumption could be tied with high power appliances such as pool pump, multiple AC units, and simultaneous usage of wet appliance or electric vehicle (EV) charging.

In another category of distributed energy management applications, the increased integration of renewable resources such as solar generation is of utmost importance to move towards energy decarbonization. Clusters 9, 16, 26, and 35, whose demand patterns in the noon timeframe coincide with the solar generation, could benefit more from photovoltaic (PV) integration. Clusters 20 and 29 have almost zero demand at noon, implying they can store considerable energy if they have PV-battery systems. Alternatively, with the rise of peer-to-peer energy trading between prosumers and consumers, these PV-equipped houses clusters could offer their high amount of on-site generation to their consumer neighbors. Finally, clusters 14, 15, 26, and 27 have sharp peaks after midnight, probably due to EV charging and wet appliances operation. Since heavy usage time is during off-peak demand, their energy behavior is compatible with DR plans, assuming the typical evening/night network peak demand.
\section{Conclusion}
\label{conclusion}
The wide availability of smart meter infrastructure and energy time-series data provides opportunities for customers’ energy behavior analytics. Clustering energy load shapes into a small number of representative patterns helps utilities in resource allocation and energy-efficiency program design. However, clustering using cluster validation indices (CVIs) may result in oversimplified load shape representations or clusters whose centroids deviate from their associated load shapes. We introduced a two-stage clustering method to preserve temporal patterns and peak magnitude of load shapes for segmentation. We first provided a comparative assessment of conventional clustering techniques and CVIs. We then showed that using common CVIs underestimates the number of clusters. The proposed approach utilized overpopulation and merging as a method to improve cluster quality, measured by correlation between cluster centroids and individual members. Empirical investigation and quantified assessment compared to the benchmark solutions were presented to demonstrate the applicability of the method.
In future, we plan to 1) investigate this approach in domains other than building science, 2) develop pre-processed cluster library for classifying energy use behavior of new customers, 3) update the library with new clusters (unseen energy behavior), and 4) investigate the correlation of time-of-use of appliances with clusters of electricity load shapes obtained by the proposed approach.
\bibliographystyle{IEEEtran}
\bibliography{main.bib}
\end{document}